\documentstyle[11pt,a4wide]{article}
\newcommand{\bean}{\begin{eqnarray*}}
\newcommand{\eean}{\end{eqnarray*}}
\newcommand{\half}{\frac{1}{2}}
\newcommand{\quarter}{\frac{1}{4}}
\newcommand{\delslash}{2\pi\delta}
\newcommand{\veck}{\vec{k}}

\newcommand{\vecx}{\vec{x}}
\newcommand{\dmuu}{\partial^{\mu}}
\newcommand{\dmul}{\partial_{\mu}}
\newcommand{\dnuu}{\partial^{\nu}}

\newcommand{\lag}{{\cal{L}}}
\newcommand{\prop}{ \Delta^{\mu\nu} }

\begin{document}
\renewcommand{\thefootnote}{\fnsymbol{footnote}}

\begin{flushright}
Imperial/TP/92-93/43 \\
12th January, 1994 \\
hep-ph/9307229
\end{flushright}

\vskip 1cm
\begin{center}
{\Large\bf Real Time Thermal Propagators for Massive Gauge Bosons}
\vskip 1.2cm
{\large\bf T.S.Evans\footnote{E-Mail: T.Evans@IC.AC.UK} \&
A.C.Pearson\footnote{E-mail: A.Pearson@IC.AC.UK}}\\
Blackett Laboratory, Imperial College, Prince Consort Road,\\
London SW7 2BZ  U.K. \\
\end{center}

\vskip 1cm
\begin{abstract}
We derive Feynman rules for gauge theories exhibiting spontaneous
symmetry breaking using the real-time formalism of finite temperature
field theory. We also derive the thermal propagators where only the
physical degrees of freedom are given thermal boundary conditions.
We analyse the abelian Higgs model and find that these new
propagators simplify the calculation of the thermal contribution to
the self energy.
\end{abstract}

\vskip 1cm
\renewcommand{\thefootnote}{\arabic{footnote}}
\setcounter{footnote}{0}

\section{Introduction} Since the observation by Kirzhnits and Linde
\cite{KL} that spontaneously broken symmetries may be restored at
high temperatures, the subject has attracted wide attention.
However, very little study has been made of theories containing
spontaneous symmetry breaking at non-zero temperature using the
so-called real time formalisms. This seems odd since in this
formalism, dynamical quantities are more readily calculated as the
thermal Green functions are given directly in terms of real times .
We are only aware of one paper, by Ueda\cite {Ueda}, that uses the
real time formalism to study the abelian Higgs model at non-zero
temperature. However, this work was done prior to the further
development of real time methods by Niemi \& Semenoff\cite{NS} and
others. We therefore feel that a study of spontaneously broken gauge
symmetries using the real time formalism is necessary.

In this paper we begin by deriving the real time thermal propagator
for a massive gauge boson in the covariant gauges. We find that
there are two thermal contributions to the propagator. The first
part is consistent with the interpretation of a field in a heat bath
of real particles. However the second term acts as a counterterm to
the finite temperature contributions from the unphysical goldstone
modes. We then apply the the ideas of Landshoff \& Rebhan\cite{LR}
and only apply the thermal boundary conditions to the physical
degrees of freedom. Previously, this method has only been applied to
gauge theories where the full symmetry is maintained.

In the final section , we use the abelian Higgs model as a simple
example. We analyse the calculation of the one-loop corrections to
the mass of the Higgs boson performed by Ueda\cite{Ueda}. While we
do not disagree with the the results therein, there were a number of
mistakes in the form of the propagators used. We evaluate the
propagators correctly. We also show that if we use propagators for which
only the physical modes are thermalised, the calculation is
greatly simplified.

\section{Derivation of RTF Propagator}

To begin with let us consider a set of gauge fields $A^{\mu}_{a}$
invariant under some gauge group. After symmetry breaking , the
free field lagrangian for these fields is of the form;

\begin{equation}
\lag_{0} = -\quarter \big{(} \dmuu A^{\nu}_{a} - \dnuu A^{\mu}_{a}
\big{)}^{2} + \half M^{2}_{ab} A^{\mu}_{a} A_{\mu b} +
\frac{1}{\zeta} \bigg{(} \dmuu A^{a}_{\mu} \bigg{)}^{2} \label{alag}
\end{equation}

Here we have chosen, like Ueda\cite{Ueda} to work in the covariant
gauges $\dmul A^{\mu}_{a} = constant$. In general we have to
include the interactions between the gauge fields and Fadeev-Popov
ghosts. However there are no problems that arise that have not
already been considered by Kobes, Semenoff \& Weiss\cite{KSW}.

\begin{figure}[htb]

\begin{center}
\setlength{\unitlength}{.3in}
\begin{picture}(5,5.5)

\thicklines

%  *** axes ***

\put(-0.5,4.6){\vector(1,0){6.0}}
\put(5.0,4.7){\large $\Re e \, (t) $ }
\put(2.5,0.0){\vector(0,1){5.0}}
\put(2.6,5.0){\large $\Im m \, (t) $ }

\thinlines

% *** RTF curve ***

\put(0,4.5){\vector(1,0){2.5}}
\put(2.5,4.5){\line(1,0){2.5}}

\put(5,4.5){\vector(0,-1){1}}
\put(5,3.5){\line(0,-1){1}}
\put(5,2.5){\vector(-1,0){2.5}}
\put(2.5,2.5){\line(-1,0){2.5}}

\put(0,2.5){\vector(0,-1){1}}
\put(0,1.5){\line(0,-1){1}}

\put(1.6,2.9){\makebox(0,0){\large $-\imath \half  \beta$}}
\put(1.7,0.5){\makebox(0,0){\large $-\imath \beta$}}
\put(3,4){${\cal C}_{1}$}
\put(3,1.8){${\cal C}_{2}$}
\put(5.1,3.3){${\cal C}_{3}$}
\put(0.2,1.5){${\cal C}_{4}$}

\end{picture}

\end{center}

\caption{Path used for the Real Time Formalism.}
\label{fig:RTF}

\end{figure}
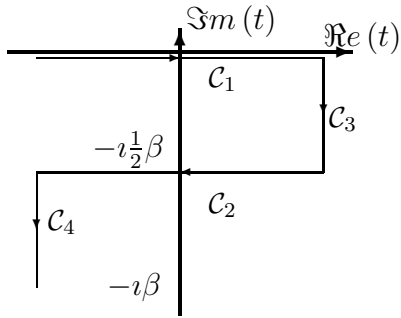

The mass matrix $M^{2}_{ab}$ depends on the particular pattern of
symmetry breaking. However, since the mass matrix is symmetric we
are able to diagonalise $M^{2}_{ab}$ by an orthogonal
transformation of the gauge fields. Once the mass matrix is
diagonalised, the gauge fields decouple in Eqn.(\ref{alag}) and we
may treat them seperately. We need therefore only consider the case
of a single gauge field.  In this case, we find that the propagator
satisfies the equation

\begin{equation}
\bigg{[} ( \Box + M^{2} ) g^{\mu\nu} + ( \frac{1}{\zeta} - 1 ) \dmuu
\dnuu \bigg{]} \Delta_{\nu\rho} (t-t') = g^{\mu}_{\rho}
\delta_{C}(t-t') \label{wave}
\end{equation}
subject to the KMS condition\cite{KMS}.
The subscript $C$ denotes that we are considering the solution to
this equation along the contour in the complex time plane associated
with the real time formalism\cite{NS,RealT,BANFF,TNC} (see fig.\ref{fig:RTF}).

We take as an Ansatz $ \prop = g^{\mu\nu} A(\vecx,t) +
\dmuu\dnuu B(\vecx,t) $.
The solution of (\ref{wave}) is
\begin{equation}
\prop = g^{\mu\nu} A(\vec{x},t;M^{2}) + \frac{\dmuu\dnuu}{M^{2}}
\bigg{(} A(\vec{x},t;M^{2}) - A(\vec{x},t;\zeta M^{2}) \bigg{)}
\label{Spa}
\end{equation}
where
\[
A(\vec{x} ,t;M^{2}) = \int d^{3}{\bar{k}} e^{\imath\veck.\vecx}
\frac{\imath}{2\omega} \frac{1}{\exp(\beta\omega)-1}
\bigg{(} ( e^{-\imath\omega t} + e^{-\beta\omega} e^{\imath\omega t}
) \Theta (t) + ( e^{\imath\omega t} + e^{-\beta\omega} e^{-\imath\omega t}
) \Theta (-t) \bigg{)}
\]
is the solution of $\big{(} \Box + M^{2} \big{)} A(\vec{x},t-t';M^{2})=
\delta_{C}(t-t')$, $\omega^{2}=\veck^{2}+M^{2}$ and
$\frac{\partial\Theta(t-t')}{\partial t} =
\delta_{c}(t-t')$.

When using the RTF of finite temperature field theory, we have ghost
fields dual to each field of the theory.
Vertices couple physical or ghost fields only to themselves, the
coupling constants differing by a minus sign.
Physical and ghost fields only mix through the propagator which now
becomes a $2 \times 2$ matrix dependent only on real times.
The components of this matrix are defined by
\begin{eqnarray}
\Delta_{ij}(t,t') = \Delta(t-t') & t\in C_{i}, t'\in C_{j} & i,j=1,2
\end{eqnarray}
For a more complete derivation of this see \cite{NS,RealT,BANFF,TNC}.

Using this result, we may now derive the real-time propagator for the
massive photon at finite temperature.
In Fourier space, the propagator is
\begin{eqnarray}
\prop(k) & = & -\bigg{(} g^{\mu\nu} - \frac{k^{\mu}k^{\nu}}{M^{2}}
\bigg{)} B
\left[ \begin{array}{cc}
    \Delta(k) & 0 \\
    0 & -\Delta^{*}(k)
\end{array} \right]
B \nonumber \\
& & - \frac{k^{\mu}k^{\nu}}{M^{2}} B
\left[ \begin{array}{cc}
    \Gamma(k) & 0 \\
    0 & -\Gamma^{*}(k)
\end{array} \right]
B \label{Answer}
\end{eqnarray}
where
\[
\begin{array}{cclccl}
\Delta(k) & = & \frac{1}{k^{2}-M^{2}+\imath\varepsilon}, &\Gamma(k)
& = & \frac{1}{k^{2}-\zeta M^{2}+\imath\varepsilon} \\
& \\
B &=& \left[
\begin{array}{ll}
\cosh\theta & \sinh\theta \\
\sinh\theta & \cosh\theta
\end{array}
\right],
& \sinh^{2}\theta & = & \frac{1}{e^{\beta | k_{0} |}-1}
\end{array}
\]
Consider the $1$-$1$ component of this propagator which is interpreted
as the propagator between the physical fields.
Explicitly the $1$-$1$ component is
\begin{eqnarray}
\imath\prop_{11}(k) & = & \frac{-\imath}{k^{2}-M^{2}+\imath\varepsilon}
\bigg{(} g^{\mu\nu} - \frac{(1-\zeta)k^{\mu}k^{\nu}}{k^{2}-\zeta M^{2}
+ \imath\varepsilon} \bigg{)} \nonumber \\
& & - \bigg{(} g^{\mu\nu} - \frac{k^{\mu}k^{\nu}}{M^{2}} \bigg{)}
\bigg{[} \frac{\delslash(k^{2}-M^{2})}{e^{\beta |k_{0}|}-1} \bigg{]}
- \frac{k^{\mu}k^{\nu}}{M^{2}} \bigg{[}
\frac{\delslash(k^{2}-\zeta M^{2})}{e^{\beta |k_{0}|}-1} \bigg{]}
\label{Grail}
\end{eqnarray}
It can be seen that the propgator splits into the $T=0$ and $T>0$
contributions.
in the limit as $T \rightarrow 0$ ($\beta \rightarrow \infty$) the $T
> 0$ terms disappear and $\prop_{11}$ reduces to the zero temperature
propagator.

We note that there are two finite temperature contributions to the
form of this propagator. The first term containing
`$\delta(k^{2}-M^{2})$' may be interpreted as a contribution due to
a heat bath of real particles of mass $M$. However, the second term
contains the unphysical gauge parameter $\zeta$ and its
interpretation is not so straightforward. The second term acts as
a counterterm to the unphysical contributions from the goldstone
boson. This can be shown in two ways. Firstly, let us consider this
model in the unitary gauge. In this gauge, the goldstone boson is
gauged away and so we would expect there to be no counterterms in
the photon propagator. The unitary gauge propagator satisfies the
following equation.
\begin{equation}
\bigg{[} ( \Box + M^{2} ) g^{\mu\nu} - \dmuu \dnuu \bigg{]}
\Delta_{\nu\rho} (t-t') = g^{\mu}_{\rho} \delta_{C}(t-t')
\end{equation}
Evaluating this equation we find
\begin{equation}
\prop(k) = - \bigg{\{} g^{\mu\nu} - \frac{k^{\mu} k^{\nu}}{M^{2}}
\bigg{\}} B \left[
\begin{array}{cc}
\Delta(k) & 0 \\
0 & -\Delta^{*}(k)
\end{array}
\right] B
\end{equation}
By comparison with Eqn.(\ref{Answer}), it can be seen that the term
dependent on the unphysical gauge parameter is not present.

As an alternative way of seeing that the covariant gauge photon
propagator contains an unphysical pole is to use the idea of
Landshoff \& Rebhan\cite{LR} and apply the KMS condition only to the
physical modes of the theory. The unphysical modes will keep the
zero temperature boundary conditions. Here we apply this method to
spontaneously broken gauge theories. Using these conditions the
goldstone boson, being unphysical has only the ero temperature
propagator. The photon propagator becomes

\begin{eqnarray}
\prop(k) & = & -\bigg{(} g^{\mu\nu} - \frac{k^{\mu}k^{\nu}}{M^{2}}
\bigg{)} B
\left[ \begin{array}{cc}
    \Delta(k) & 0 \\
    0 & -\Delta^{*}(k)
\end{array} \right]
B \nonumber \\
& & - \frac{k^{\mu}k^{\nu}}{M^{2}}
\left[ \begin{array}{cc}
    \Gamma(k) & 0 \\
    0 & -\Gamma^{*}(k)
\end{array} \right]
\end{eqnarray}
The 1-1 component of this propagator is
\[
\prop(k)= \frac{-\imath}{k^{2}-M^{2}+\imath\varepsilon} \bigg{(}
g^{\mu\nu}- \frac{(1-\zeta) k^{\mu} k^{\nu}}{k^{2}-\zeta M^{2} +
\imath \varepsilon} \bigg{)} - \bigg{(} g^{\mu\nu}- \frac{k^{\mu}
k^{\nu}}{M^{2}} \bigg{)}
\bigg{[}\frac{2\pi\delta(k^{2}-M^{2})}{e^{\beta |k_{0}|}-1} \bigg{]}
\]
Comparing this result with Eqn.(\ref{Answer}), one can see that
although we have retained the usual gauge dependent, zero
temperature contribution to the propagator, the gauge dependent
finite temperature contribution is not present.

Finally let us consider the limit as the mass of the gauge boson
vanishes. This case is problematic as the propagator contains terms
proportional to $M^{-2}$. To proceed we must return to
Eqn.(\ref{Spa}) and examine the massless limit. Using
L'H\^{o}pital's rule, Eqn.(\ref{Spa}) becomes
\begin{equation}
\prop = g^{\mu\nu} A(\vecx,t;M^{2}=0) + (1-\zeta)\dmuu\dnuu
\frac{\partial A(\vecx,t;M^{2})}{\partial M^{2}}\bigg{|}_{M^{2}=0}
\label{Ans}
\end{equation}
In momentum space this becomes,
\begin{equation}
\prop(k) = -\bigg{(} g^{\mu\nu} - k^{\mu}k^{\nu}
\frac{\partial}{\partial k^{2}} \bigg{)} B
\left[ \begin{array}{cc}
    \Delta(k) & 0 \\
    0 & -\Delta^{*}(k)
\end{array} \right] B
\end{equation}
where $\Delta(k)=\frac{\imath}{k^{2}+\imath\varepsilon}$.
This is exactly the propagator for a photon first derived by Kobes,
Semenoff, and Weiss in \cite{KSW}.

\section{The abelian Higgs model}

The propagator we have derived for the massive gauge boson differs
from the result used by Ueda (Eqn.3.8, \cite{Ueda}). However this work was done
prior to the formulation of real time thermal field theory\cite{RealT} using
path integral methods devloped by Niemi \& Semenoff\cite{NS}. We
would therefore like to reconsider the abelian Higgs model.

The abelian Higgs model consists of a $U(1)$ gauge field coupled to
a complex scalar field described by the Lagrangian density
\[
\lag = -\quarter F^{\mu\nu} F_{\mu\nu} + \big{(} D_{\mu} \Phi
\big{)}^{*} \big{(} D^{\mu} \Phi \big{)} - \rho^{2} \Phi^{*} \Phi -
\frac{\lambda}{6} \big{(}\Phi^{*}\Phi\big{)}^{2}
\]
where $F^{\mu\nu}=\dmuu A^{\nu} \dnuu A^{\mu}$ and $D^{\mu}=
\dmuu-\imath e A^{\mu}$. We choose $\rho^{2} < 0$ so that
spontaneous symmetry breaking occurs.

Expanding $\Phi$ about its expectation value as
$\Phi(x)=\frac{1}{\surd{2}} \big{(}v+\phi(x)+ \imath\chi(x)\big{(}$
where $\langle\Phi(x)\rangle=\frac{v}{\surd{2}}$, we may rewrite the
lagrangian as $\lag= \lag_{0} -V(A^{\mu},\phi ,\chi)$ where
\begin{eqnarray}
\lag_{0} &= & -\quarter F^{\mu\nu} F_{\mu\nu} + \half M^{2} A^{\mu}
A_{\mu} - \frac{1}{2\zeta} \big{(} \dmul A^{\mu} \big{)}^{2}
\nonumber \\
& & + \half\dmuu\phi\dmul\phi -\half m^{2}\phi^{2} +
\half \dmuu\chi\dmul\chi - M\dmul\chi A^{\mu} \label{free}
\end{eqnarray}

and $V(A^{\mu},\phi ,\chi )$ is a potential term containing only
cubic and quartic terms. $M= ev$ and $m^{2}=\rho^{2}+ \half\lambda
v^{2}$.
We have added a gauge fixing term to $\lag_{0}$. We choose to work
in the covariant gauges $\dmul A^{\mu}=$constant. Using
Eqn.\ref{free}, we find that the 1-1 components of the thermal
propagators are
\begin{eqnarray}
\imath\Delta_{H} &= &\frac{\imath}{k^{2}-m^{2}+\imath\varepsilon} +
\frac{\delslash(k^{2}-m^{2})}{e^{\beta |k_{0}|}-1} \nonumber \\
\nonumber\\
\imath\Delta_{G} &= & \frac{\imath}{k^{2}+\imath\varepsilon} +
\frac{\delslash(k^{2})}{e^{\beta |k_{0}|}-1} \nonumber \\
\nonumber\\
\imath\prop(k) & = & \frac{-\imath}{k^{2}-M^{2}+\imath\varepsilon}
\bigg{(} g^{\mu\nu} - \frac{(1-\zeta)k^{\mu}k^{\nu}}{k^{2}-\zeta
M^{2} + \imath\varepsilon} \bigg{)} \nonumber \\  & & - \bigg{(}
g^{\mu\nu} - \frac{k^{\mu}k^{\nu}}{M^{2}} \bigg{)} \bigg{[}
\frac{\delslash(k^{2}-M^{2})}{e^{\beta |k_{0}|}-1} \bigg{]} -
\frac{k^{\mu}k^{\nu}}{M^{2}} \bigg{[} \frac{\delslash(k^{2}-\zeta
M^{2})}{e^{\beta |k_{0}|}-1} \bigg{]}
\end{eqnarray}
Before one can use these propagators, one must take into account the
effect of the $A_{\mu}$-$\chi$ mixing term in Eqn.(\ref{free}). The
effect of this term is to modify the photon and goldstone
propagators and to introduce a new photon-goldstone vertex.
Explicitly, these modified propagators are
\begin{eqnarray}
\imath\Delta^{mod}_{G} &= &\bigg{[} 1+ \zeta M^{2}
\frac{\partial}{\partial k^{2}} \bigg{]} \bigg{(}
\frac{\imath}{k^{2}+\imath\varepsilon} +
\frac{\delslash(k^{2})}{e^{\beta | k_{0} |}-1} \bigg{)}
\nonumber \\
\imath\prop_{mod}(k) &=
&-\bigg{\{} \frac{\imath}{k^{2}-M^{2}+\imath \varepsilon} +
\frac{\delslash(k^{2}-M^{2})}{e^{\beta | k_{0} |}-1} \bigg{\}}
\bigg{(} \frac{\zeta k^{\mu} k^{\nu}}{k^{2}}g^{\mu\nu} -
\frac{k^{\mu}k^{\nu}}{k^{2}} \bigg{)}
\nonumber \\
& & -\frac{k^{\mu} k^{\nu} }{M^{2}}
\frac{\delslash(k^{2})}{e^{\beta | k_{0} |}-1}
+ \imath\zeta k^{\mu} k^{\nu} \frac{\partial}{\partial k^{2}}
\bigg{[} \frac{\imath}{k^{2}+\imath\varepsilon} +
\frac{\delslash(k^{2})}{e^{\beta | k_{0} | }-1} \bigg{]}
\end{eqnarray}

Comparing this with Eq.3.12 of \cite{Ueda}, we can see that although
the bare propagator used by Ueda was incorrect, the modified
propagator has is correct form except for the gauge dependent term.
However, the only problem with this final term was, as noted by
Ueda, the problem of understanding terms such as
$\frac{\delta(k^{2})}{k^{2}}$. As it was pointed out by Kobes,
Semenoff \& Weiss\cite{KSW}, these terms should instead be
considered as derivatives of the delta function. As we find that the
modified propagator used by Ueda is essentially correct, it is no
surprise that we agree with the result of the one-loop self energy
corrections calculated in \cite{Ueda} and we shall not repeat the
calculation here. However, we would like to consider this
calculation using propagators for which only the physical degrees of
freedom are given thermal boundary conditions\cite{LR}. Using these
propagators the modified propagators due to photon-goldstone mixing
are

\begin{eqnarray}
\imath\prop_{mod}(k) &=&-\bigg{\{} \frac{\imath}{k^{2}-M^{2}+\imath
\varepsilon} + \frac{\delslash(k^{2}-M^{2})}{e^{\beta | k_{0} |}-1}
\bigg{\}} \bigg{(} g^{\mu\nu} - \frac{k^{\mu}k^{\nu}}{k^{2}} \bigg{)}
-\frac{\imath}{k^{2}+\imath\varepsilon} \frac{\zeta k^{\mu}
k^{\nu}}{k^{2}} \nonumber \\
\nonumber\\
\imath\Delta_{G}^{mod} &=& \bigg{[} 1-\frac{\zeta M^{2}}{k^{2}}
\bigg{]} \frac{\imath}{k^{2}+\imath\varepsilon}
\end{eqnarray}
It can be seen the gauge dependent terms are in their zero
temperature form as expected. As such when calculating the one-loop
thermal contribution to the self energy, the number of gauge
dependent terms are greatly reduced. Using the original propagators
there are 12 different diagrams which have gauge and temperature
dependent terms whereas using these new propagators there are just
3. However, we still retain the gauge parameter in our calculations
providing a useful check that we have correctly evaluated the
diagrams.

Since we are only applying thermal boundary conditions to the
physical modes of the model, the thermal contributions to these
modified propagators are identical to the thermal contributions found
when using the unitary gauge. One might therefore be worried that
the unitary gauge puzzle\cite{UGP} might appear when using these
propagators. However, this is not the case. The unphysical thermal
contributions do not appear in our propagators because of our choice
of boundary conditions, not because of the choice of gauge.

\section{Conclusions}

In this paper, we have derived the real time thermal propagators for
massive gauge bosons. We have seen that the thermal contribution to
the propagator contains gauge dependent terms. These terms are
unphysical as can be seen by deriving the thermal propagator in the
unitary gauge. Also we have studied the propagator when thermal
boundary conditions are applied to the physical degrees of freedom
and zero temperature boundary conditions to any unphysical modes.
Using these propagators we find that the calculation of the thermal
contribution to the one-loop self energy of the Higgs particle is
greatly simplified. In fact, the number of diagrams to be calculated
is to be compared with the same calculation performed in the unitary
gauge. However, we retain the gauge parameter and avoid any of the
problems associated with the unitary gauge puzzle\cite{UGP}.

\pagebreak

\end{document}